%% file: Verification_of_Concurrent_Binary_Search_Tree.tex
\documentclass[runningheads]{llncs}

\usepackage{txfonts}
\usepackage[T1]{fontenc}
\usepackage{graphicx}
\usepackage{subcaption} 
\usepackage{uri}
\usepackage{mathpartir}
\usepackage{semantic}
\usepackage{cases}
\usepackage{hyperref}
\usepackage{color}

\usepackage{stmaryrd}
\usepackage{listings}
\usepackage{amssymb}
\usepackage{iris-llncs}
\usepackage[edges]{forest}

\newcommand{\islock}{\boxdotright}
\newcommand{\lockvar}{\islock^{v}}

\newcommand\dboxed[1]{\dbox{\ensuremath{#1}}}

\newcommand{\ghost}[2]{\ensuremath{\dboxed{#1}^{#2}}}

\newcommand{\gnamety}{\ensuremath{\mathsf{gname}}}
\newcommand{\treerep}{\ensuremath{\mathsf{bst\_abs}}}
\newcommand{\nodeboxrep}{\ensuremath{\mathsf{bst\_ref}}}

\newcommand{\nodeghost}[2]{\ensuremath{\mathsf{node\_ghost}_{#1}\ (#2)}}
\newcommand{\myhalf}[2]{\ensuremath{\mathsf{node\_ghost}^{.5}_{#1}\ (#2)}}
\newcommand{\publichalf}[2]{\ensuremath{\mathsf{node\_ghost}^{.5}_{#1}\ (#2)}}

\newcommand{\ignore}[1]{}

\usepackage[utf8]{inputenc}

\begin{document}


\title{Proving Logical Atomicity using Lock Invariants\thanks{Supported by the National Science Foundation via grant \#1811894.}}
%
%
\author{Roshan Sharma\inst{1} \and
Shengyi Wang\inst{2}\orcidID{0000-0002-2286-8703} \and
Alexander Oey\inst{3} \and
Anastasiia Evdokimova\inst{4} \and
Lennart Beringer\inst{2}\orcidID{0000-0002-1570-3492} \and
William Mansky\inst{4}\orcidID{0000-0002-5351-895X}} 

\authorrunning{R. Sharma et al.}

\institute{Amazon, USA \and Princeton University, USA \and Cisco, USA \and University of Illinois Chicago, USA} 



\maketitle

\begin{abstract}
Logical atomicity has been widely accepted as a specification format for data structures in concurrent separation logic. While both lock-free and lock-based data structures have been verified against logically atomic specifications, most of the latter start with atomic specifications for the locks as well. In this paper, we compare this approach with one based on older lock-invariant-based specifications for locks. We show that we can still prove logically atomic specifications for data structures with fine-grained locking using these older specs, but the proofs are significantly more complicated than those that use atomic lock specifications. Our proof technique is implemented in the Verified Software Toolchain, which relies on older lock specifications for its soundness proof, and applied to C implementations of lock-based concurrent data structures.

\end{abstract}


\section{Introduction}
\input{intro.tex}

\section{Background}
\label{background}

\subsection{Logically Atomic Specifications}
\label{atomic-triples}

For many concurrent data structures, the ideal correctness condition is that the data structure behaves the same as a sequential implementation, even when accessed simultaneously by multiple threads. This intuitive condition has been formalized in terms of linearizability (operations appear to take effect in some total order) or atomicity (client threads never observe intermediate states of operations). In separation logic, atomicity can be expressed in the form of \emph{atomic triples}~\cite{tada} of the form \[\langle a.\ P_l\ |\ P_p(a)\rangle\ c\ \langle Q_l\ |\ Q_p(a)\rangle\] where $P_l$ and $Q_l$ are \emph{local} pre- and postconditions similar to an ordinary Hoare triple, and $P_p$ and $Q_p$ are \emph{public} pre- and postconditions, parameterized by an abstract value $a$ of the shared data structure. This triple says that if $P_l$ is true before a call, and the fact that $P_p$ is true for some value of $a$ is held in shared state (e.g. a global invariant), then $P_p$ will continue to be true for some (possibly different) value of $a$ up until the \emph{linearization point} of $c$, at which point $Q_p$ will become true atomically for the same value $a$ (and $Q_l$ will be true after $c$ ends). 
The specific value of $a$ (the shared abstract state) can vary as other threads modify the data structure, as long as $P_p(a)$ is always maintained. For instance, the specification \[\langle s.\ \mathsf{is\_stack}\ p\ |\ \mathsf{stack}\ s\rangle\ \texttt{push}(v)\ \langle \mathsf{is\_stack}\ p\ |\ \mathsf{stack}\ (v :: s)\rangle\] expresses the fact that the \texttt{push} operation of a concurrent stack correctly implements the behavior of a sequential push, making no visible changes to whatever stack $s$ is in place until, at the linearization point, the current stack $s$ is replaced atomically with $v :: s$. The $\mathsf{stack}$ itself is treated as a publicly owned resource, and can only be accessed and modified atomically by threads holding the $\mathsf{is\_stack}$ assertion.

\subsection{Two Styles of Lock Specification}
Soon after the introduction of concurrent separation logic~\cite{csl}, both Gotsman et al.~\cite{gotsman} and Hobor et al.~\cite{oraclesematic} extended it with support for C-style storable locks, where a lock is a special memory location designated as coordinating access to some other resource. Their lock specifications were essentially equivalent, and of the form:
\[\{ \ell \islock R \}\ \texttt{acquire}(\ell)\ \{ R * \ell \islock R \} \qquad \{ R * \ell \islock R \}\ \texttt{release}(\ell)\ \{ \ell \islock R \}\]
Each lock $\ell$ is associated with a ``lock invariant'' $R$, an arbitrary separation logic predicate representing the protected resource, which is gained by a thread that acquires the lock and must be restored when the lock is released. At the time, there was no formal treatment of atomic operations in CSL, so actual lock implementations were beyond the scope of verification: these specs were taken as axioms, or proved through direct appeal to data-race-free guarantees of specific memory models. In garbage-collected languages that do not deallocate locks, the assertion $\ell \islock R$ can be freely duplicated and shared among threads accessing $R$; if we want to track ownership and eventually deallocate the lock, we can instead split the assertion into fractional shares such that $\ell \islock_{\pi_1} R * \ell \islock_{\pi_2} R = \ell \islock_{\pi_1 + \pi_2} R$, and then allow deallocation only when we own the full share $\ell \islock_1 R$.

Later, the TaDA logic~\cite{tada}, which codified the notion of logical atomicity, presented an alternative set of lock specifications:
\[\langle b.\ (\mathsf{L}(\ell) \land \neg b) \vee (\mathsf{U}(\ell) \land b) \rangle\ \texttt{acquire}(\ell)\ \langle \mathsf{L}(\ell) \land b \rangle \qquad \langle \mathsf{L}(\ell) \rangle\ \texttt{release}(\ell)\ \langle \mathsf{U}(\ell) \rangle\]
An \texttt{acquire} operation may access the lock repeatedly, finding it in either the locked ($\mathsf{L}$) or unlocked ($\mathsf{U}$) state; if it sees the unlocked state ($b$ holds), then it can atomically move to a locked state and return. A release simply moves atomically from the locked state to the unlocked state. These operations do not specify what is protected by the lock, but because they are atomic, they can interact with any global invariant (``shared region'' in TaDA's terminology): in particular, in combination with the invariant $(\mathsf{U}(\ell) * R) \vee \mathsf{L}(\ell)$, they can be used to derive the lock-invariant-based specs. It is also worth noting that these lock specifications are not axioms: they are proved for a simple spinlock implementation that uses atomic operations (e.g. compare-and-swap), which are the actual primitive operations in TaDA and the logics that follow it. In this style, tracking ownership of a lock is orthogonal to the lock assertion itself; a mechanism such as cancelable invariants~\cite{rustbelt-relaxed} can be used to both make the lock assertion $\mathsf{U}$/$\mathsf{L}$ publicly available and deallocate it once all threads are finished.

\subsection{VST and Iris}

We employ a combination of two Coq-based verification systems, the Verified Software Toolchain (VST) and Iris. VST~\cite{plfcc} is a Coq-based system for proving separation logic specifications of C programs. It is proved sound against not only the C semantics of the CompCert compiler, but also the compiler correctness theorem~\cite{cpm}, so that verified programs are guaranteed to behave correctly when compiled and run. VST directly implements the concurrent separation logic of Hobor et al.~\cite{oraclesematic}, and the invariant-style lock specifications are incorporated into its soundness proof at a fundamental level.

Iris~\cite{iris} is a framework that synthesizes ideas from a range of CSLs with the idea that ``ghost state is all you need''. Iris begins with a small core logic including arbitrary higher-order ghost state, and derives features including invariants and logical atomicity from particular instances of ghost state. Iris is designed as a language- and logic-independent framework, and recent versions of VST take advantage of this, incorporating Iris-style ghost state in the foundational model and directly using Iris tactics to reason about invariants and atomicity of C implementations. All proofs described in this paper have been formalized in this VST+Iris setting. Iris has also been used as the basis for proofs with atomic lock specifications, such as the work of Krishna et al.~\cite{templates}.

\section{Locks and Atomicity}
\label{atomic}
In this section, we outline our approach to connecting lock invariants and atomic specifications, and contrast it with the atomic-lock-based approach.

Consider a sequential implementation of a binary search tree. It has functions \texttt{insert}, \texttt{lookup}, and \texttt{delete}, proven to satisfy the following specifications:
\begin{mathpar}
\{\mathsf{BST}\:p\ t\}\ \texttt{insert}(p, k, v)\ \{\mathsf{BST}\:p\ (t[k \mapsto v])\}

\{\mathsf{BST}\:p\ t\}\ \texttt{lookup}(p, k)\ \{v.\ \mathsf{BST}\:p\ t \land t(k) = v\}

\{\mathsf{BST}\:p\ t\}\ \texttt{delete}(p, k)\ \{\mathsf{BST}\:p\ (t[k \mapsto \_])\}
\end{mathpar}
where $\mathsf{BST}\:p\ t$ asserts that some abstract binary search tree $t$ is represented in memory starting at location $p$. For instance, $t$ could be an element of a recursively defined tree type in Coq, and $\mathsf{BST}$ could be defined as a recursive predicate that represents each node of $t$ as an object in memory, with fields for key, value, and pointers to left and right children, with the root node located at $p$.

A thread-safe version of this data structure should satisfy atomic specifications that closely correspond to the sequential specs\footnote{We also have functions to create and deallocate trees, but these functions need not be thread-safe, since the tree will not be shared when created and must not be shared when deallocated.}:
\begin{mathpar}
\langle m.\ \nodeboxrep_g\ p\ |\ \treerep_g\ m\rangle\ \texttt{insert}(p, k, v)\ \langle \nodeboxrep_g\ p\ |\ \treerep_g\ (m[k \mapsto v])\rangle

\langle m.\ \nodeboxrep_g\ p\ |\ \treerep_g\ m\rangle\ \texttt{lookup}(p, k)\ \langle v.\ \nodeboxrep_g\ p\ |\ \treerep_g\ m \land m(k) = v\rangle

\langle m.\ \nodeboxrep_g\ p\ |\ \treerep_g\ m\rangle\ \texttt{delete}(p, k)\ \langle \nodeboxrep_g\ p\ |\ \treerep_g\ (m[k \mapsto \_])\rangle
\end{mathpar}
These specifications assert that each operation appears to execute atomically, accessing and updating the state of the data structure without exposing any intermediate states---for instance, a \texttt{lookup} should not find a partially inserted key with a value other than the one passed to its \texttt{insert} call. We split the $\mathsf{BST}$ assertion into two pieces: $\nodeboxrep$ contains the points-to and lock-invariant assertions that a thread needs in order to call a BST function, and $\treerep$ contains the ghost state that describes the overall state of the tree and is treated as an atomically accessed shared resource. They are connected by reference to a fixed but arbitrary ghost state identifier $g$, which we will generally omit. We change the abstract state of the data structure from a tree $t$ to a key-value map $m$ to reflect the fact that clients cannot observe the internal structure of the tree (which will be useful when we want to rearrange nodes during concurrent access). Our sequential implementation will almost certainly not satisfy these specifications, because its functions relied on being able to access any part of the data structure at any time: the $\mathsf{BST}$ assertion included full ownership of every location in the data structure.

To show that an implementation satisfies these atomic specifications, we must show that 1) the implementation can execute safely without owning any piece of the public assertion $\treerep$, only accessing it atomically and restoring it up until the linearization point, and 2) at some linearization point, the implementation atomically transforms the current map $m$ into one that satisfies the public postcondition. The linearization point may be a (physically or logically) atomic operation, or it may occur between steps of the program (e.g., if the current thread holds a lock on the modified section of the data structure, so other threads cannot distinguish between the original state and the modified state until the lock is released). In the following sections, we discuss the general structure of these proofs for lock-based implementations using each of the two kinds of lock specifications.

\subsection{Atomicity with lock invariants}
The easiest way to synchronize a data structure is to add a lock that
each operation acquires at the start and releases at the end. These
\emph{coarse-grained} locking implementations are easy to verify with
invariant-style lock specs: the lock's invariant can contain the whole
data structure assertion (e.g. $\mathsf{BST}\:p\ t$), so 
that a thread always has full ownership of the data structure when
performing an operation.  If we only want to prove that the operation
is thread-safe, we might choose the lock invariant $\exists t.\
\mathsf{BST}\ p\ t$, asserting that we gain full ownership of the tree
in some state $t$ every time we acquire the lock. On the other hand,
if we want to prove specifications that describe changes to the data
structure---as in the atomic specification $\langle m.\ \treerep\
m\rangle\ \texttt{insert}(p, k, v)\ \langle \treerep\ (m[k \mapsto
v])\rangle$---we need ghost state to track the changes.

Specifically, we can define a piece of ghost state $\mathsf{ghost\_bst}\ t$ that is divided between the lock invariant and the public pre- and postconditions of the functions, connecting the tree in memory to the public abstract state of the data structure. For instance, we might choose a lock invariant $R_{\mathit{cg}} \triangleq \exists t.\ \mathsf{BST}\ p\ t * \mathsf{ghost\_bst}_{.5}\ t$, and then set $\nodeboxrep\ p \triangleq p \islock_\pi R_{\mathit{cg}}$ and $\treerep\ t \triangleq \mathsf{ghost\_bst}_{.5}\ t$. 
Then we can prove the atomic triples above, where at the linearization point we access $\treerep$, combine the two halves of the ghost state and update them to reflect the new state of the concrete data structure, and then fulfill the atomic postcondition with one half of the ghost state and return the other half to the lock invariant. The atomic postcondition reflects the fact that the abstract state has been accessed or changed, while the lock invariant maintains the connection between the abstract state and the actual data structure in memory. This approach to making changes to a locked data structure visible via ghost state is due to Zhang and Jung's proofs of an atomic syncer in Iris~\cite{atomic-syncer}.

\label{atomicity}

In a \emph{fine-grained} implementation, there are multiple locks on different pieces of the data structure, so the ghost state of each lock invariant cannot be the abstract state of the entire data structure. Instead, the ghost state represents the abstract state of the locked section alone, and the abstract state of the data structure is derived from the composition of the states of the locked components. Each locked component $c$ has a piece of ghost state $\mathsf{ghost}\_c$ that is split between the lock invariant and the top-level abstract state. At the end of a critical section for $c$, we atomically access the top-level abstract state, combine the two halves of $\mathsf{ghost}\_c$, update $\mathsf{ghost}\_c$ to reflect any local changes, and then show either that the update to $c$ does not change the top-level state of the data structure, or that the update to $c$ changes the top-level state to one that satisfies the atomic postcondition for the function. For instance, when deleting a node from a binary search tree, we rotate nodes to reconstruct a valid tree; at the end of the critical section for each rotation, we show either that the values in the tree are unchanged (and the tree is still well-ordered), or that we have removed the node to be deleted. 
As we will see in section~\ref{hand-over-hand}, the lock invariant for each component must also carefully account for the ownership of both that component's lock and the locks of related nodes (e.g., child nodes in a tree).

\subsection{Atomicity with atomic locks}
Proving atomic specifications for a data structure using atomic specifications for locks is more immediate. As an example, consider Krishna et al.'s verification of a hand-over-hand locking pattern for search structures~\cite{templates}. The lock specification used\footnote{In fact, this specification is derived from the basic TaDA-style atomic lock specificaton.} is \[\langle \mathsf{inFP}(n)\ |\ \mathsf{CSS}(r, C) \rangle\ \mathtt{lockNode}\ n\ \langle \mathsf{CSS}(r, C) * \mathsf{N}(n)\rangle\] where $\mathsf{inFP}(n)$ asserts that the node $n$ is in the data structure, $\mathsf{CSS}$ is the abstract state, and $\mathsf{N}(n)$ is the resources for the node $n$ (including both points-to assertions and ghost state). The abstract state assertion $\mathsf{CSS}$ includes both global ghost state and the composition of per-node state $\mathsf{N}$ for each node that is not currently locked, so that $\mathsf{N}(n)$ can be removed from the abstract state when $n$'s lock is acquired.

Note that in this case, the local precondition $\mathsf{inFP}(n)$ is merely the knowledge that $n$ is in the data structure, and does not involve any ownership---in particular, a caller does not need a share of the lock for the root node. A thread can acquire the lock for any node $n$ in the data structure by atomically accessing $\mathsf{CSS}$, extracting the sub-assertion corresponding to $n$, and switching it to the locked state. More generally, in this approach there is no need to track ownership of the locks of individual nodes, which as we will see in section~\ref{selflock} is a source of considerable complexity in lock-invariant-based proofs.

\section{A Verified Binary Search Tree with Hand-over-Hand Locking}
\label{hand-over-hand}

We demonstrate our approach to proving logically atomic specifications with invariant-based (type 1) lock specs by verifying a binary search tree (BST) with fine-grained locking in VST, using the approach of section~\ref{atomicity} to prove atomic specifications for the BST operations while using invariant-style specs for our locks (as required by VST's soundness proof).
As described in section~\ref{atomic}, we aim to define assertions $\nodeboxrep$ (the per-thread handle to the BST) and $\treerep$ (the shared abstract state) such that the operations of the concurrent BST satisfy atomic triples along the lines of \[\langle m.\ \nodeboxrep\ p\ |\ \treerep\ m\rangle\ \texttt{insert}(p, k, v)\ \langle \nodeboxrep\ p\ |\ \treerep\ (m[k\mapsto v])\rangle\] In this section, we describe the construction of the two predicates and the proofs of the operations for a BST implementation with hand-over-hand locking, using lock-invariant-based specs for our locks.

\subsection{Hand-over-Hand Locking}
\label{selflock}

Hand-over-hand locking (also called lock coupling) is a fine-grained locking pattern for traversing a data structure in which we acquire the lock for the next node before releasing the lock for the current node, thus guaranteeing that the link between the two nodes is not removed or rearranged while we traverse it. Hand-over-hand locking is a classic example of a pattern that requires lock invariants that refer to other lock invariants, dating back to the earliest papers on ``predicates in the heap'' in CSL~\cite{gotsman}. The trick is to assign each node in the data structure a lock invariant that includes the lock assertion for its child nodes; for a linked list, for instance, we might write \begin{equation}\label{inv1}R(n) \triangleq n \mapsto (d, n', \ell') * \ell' \islock R(n')\end{equation} where $n'$ is the child of $n$ and $\ell'$ is $n'$'s lock. By acquiring $n$'s lock, we learn the lock assertion for $n'$, and can acquire its lock as needed before releasing $n$.

This invariant becomes slightly more complicated when we consider \emph{ownership} of locks, as is required if we want to eventually free the lock and return its resources. In this case the lock assertion carries a share $\pi$ (often represented as a rational number in the range $(0, 1]$), where the full share $\pi = 1$ is required to free the lock while any share is sufficient to acquire or release it. We can amend our lock invariant to \begin{equation}R(n) \triangleq n \mapsto (d, n', \ell') * \ell' \islock_\pi R(n')\end{equation} for some fixed share $\pi$, but then, what happens to the share of $\ell'$ when we release the lock on $n$? Starting from a state in which we hold $n$'s lock $\ell$, this gives us:
\[\begin{array}{l}
\{ \ell \islock_\pi R(n) * n \mapsto (d, n', \ell') * \ell' \islock_\pi R(n') \}\\
\texttt{acquire}(\ell');\\
\{ \ell \islock_\pi R(n) * n \mapsto (d, n', \ell') * \ell' \islock_\pi R(n') * n' \mapsto (d, n'', \ell'') * \ell'' \islock_\pi R(n'') \}\\
\texttt{release}(\ell);\\
\{ \ell \islock_\pi R(n) * n' \mapsto (d, n'', \ell'') * \ell'' \islock_\pi R(n'') \}
\end{array}\]
In the call to \texttt{release}, we give up resources to reestablish $\ell$'s lock invariant, including the lock assertion for $\ell'$. As a result, we are left with a share of $\ell$ (the lock for the node we left behind) and no share of $\ell'$ (the lock for the current node).

Both of these problems can be solved by making the lock \emph{recursive}, including a share of $\ell$ in $\ell$'s own lock invariant\footnote{Gotsman et al. instead use a separation assertion $\mathsf{Locked}$ that is always held by the thread that acquires a lock, and is required to free the lock: this is functionally equivalent to an extra share of the lock assertion.}. We split ownership of the lock $\ell$ for a node $n$ into two shares: $\pi_1$ is held by $\ell$ itself, and $\pi_2$ is held by the lock of $n$'s parent. Then we amend definition \ref{inv1} to
\[R(n) \triangleq \ell \islock_{\pi_1} R(n) * n \mapsto (d, n', \ell') * \ell' \islock_{\pi_2} R(n')\]
Now the same sequence of operations gives us:
\[\begin{array}{l}
\{ \ell \islock_{\pi_1} R(n) * n \mapsto (d, n', \ell') * \ell' \islock_{\pi_2} R(n') \}\\
\texttt{acquire}(\ell');\\
\{ \ell \islock_{\pi_1} R(n) * n \mapsto (d, n', \ell') * \ell' \islock_{\pi_2} R(n')\ * \\\ \ell' \islock_{\pi_1} R(n') * n' \mapsto (d, n'', \ell'') * \ell'' \islock_{\pi_2} R(n'') \}\\
\texttt{release}(\ell);\\
\{ \ell' \islock_{\pi_1} R(n') * n' \mapsto (d, n'', \ell'') * \ell'' \islock_{\pi_2} R(n'') \}
\end{array}\]
When we acquire $\ell'$, we hold both $\pi_1$ and $\pi_2$ of its lock assertion; when we release $\ell$, we return share $\pi_1$ of $\ell$ along with share $\pi_2$ of $\ell'$, but retain share $\pi_1$ of $\ell'$, so we can release it later. The mechanism for defining a lock invariant that includes a share of its own lock assertion may vary across separation logics (VST provides a $\mathsf{selflock}$ assertion for exactly this purpose), but as long as such a mechanism exists, this pattern can be used to implement hand-over-hand lock invariants with share accounting, allowing us to traverse a data structure built with this pattern and then reclaim all of its resources once all threads are finished with it.

This pattern is the main source of complexity in our proofs that is not present in proofs with atomic lock specifications. For instance, in the hand-over-hand locking proofs of Krishna et al.~\cite{templates}, the only information a node holds about its children is the fact that they are in the data structure; their lock assertions are held in the global abstract state along with the rest of their resources. This significantly simplifies the specification and proof of the hand-over-hand synchronization.

\subsection{Specification of the hand-over-hand BST}
The C implementation of a node in our BST is:
\begin{lstlisting}[language = C,numbers = none]
typedef struct tree {int key; void *value;
                         struct tree_t *left, *right;} tree;
typedef struct tree_t {tree *t; lock_t *lock;} tree_t;
\end{lstlisting}
Each node (of type \lstinline{tree_t}) has a \lstinline{lock} field that holds the lock protecting the node, and a \lstinline{t} field that is either \lstinline{NULL} (for a leaf, which contains no key or value) or points to a \lstinline{tree} struct containing the node's key, value, and child pointers. To verify operations on this structure, we need to relate a \lstinline{tree_t} in memory to an abstract state (a mathematical map from keys to values), and divide ownership of those resources between the assertions $\nodeboxrep$ (the resources held by each client thread) and $\treerep$ (the abstract state of the data structure). As described in section \ref{atomicity}, our approach is to associate a piece of ghost state with each node in the tree, which will be shared between that node's lock and the abstract state in $\treerep$, while $\nodeboxrep$ will be a a pointer to the lock of the root node. We also need to make sure that our lock invariants support the hand-over-hand pattern.

\subsubsection{Key Ranges}
When we traverse the BST looking for a key $k$, we will reach either a node containing $k$, or an empty node where $k$ would appear if it was in the tree. To prove correctness of this procedure, we need ghost state that tracks where each key ``should appear'' in the current tree. As observed by Krishna et al.~\cite{krishna2017flow}, this can be done by associating each node with a lower and upper bound on the keys appearing in the subtree rooted at that node. We refer to the product of these bounds as a \emph{range}. The range of a node is inherited from its parent, based on the parent's key: if node $n$ has range $(l, r)$ and key $k$, then its left child has range $(l, k)$ and its right child has range $(k, r)$. Figure~\ref{range_bst} shows an example of a BST with each node labeled with its range, starting with $(-\infty, +\infty)$ at the root and propagated to the empty leaf nodes. 

\usetikzlibrary{positioning}

\begin{figure}[htb]
\centering
\begin{forest}
for tree={
  calign=fixed edge angles,
  calign angle=65,
  circle, draw, minimum size=3ex, inner sep=1pt, l sep=0, l=0,
        }
    [ 20, tikz={\node[right=0pt of .north east, scale=0.75]  {$(-\infty,\infty)$};}
        [15, tikz={\node[left=0pt of .north west, scale=0.75]  {$(-\infty,20)$};}]
        [30, tikz={\node[right=0pt of .north east, scale=0.75]  {$(20,\infty)$};}
            [25, tikz={\node[left=-2pt of .west, scale=0.75]  {$(20,30)$};}]
            [50, tikz={\node[right=0pt of .north east, scale=0.75]  {$(30,\infty)$};}
                [45, tikz={\node[right=0pt of .south east, scale=0.75]  {$(30,50)$};}
                    [40, tikz={\node[left=0pt of .north west, scale=0.75]  {$(30,45)$};}
                        [35, tikz={\node[left=0pt of .north west, scale=0.75]  {$(30,40)$};}
                            [, tikz={\node[left=0pt of .north west, scale=0.75]  {$(30,35)$};}]
                            [, tikz={\node[right=0pt of .north east, scale=0.75]  {$(35,40)$};}]
                        ]
                        [,no edge, draw=none]
                    ]
                    [,no edge, draw=none]
                ]
                [,no edge, draw=none]
            ]
        ]
    ]
\end{forest}
\caption{A BST with each node labeled with its range; empty children of nodes other than 35 are omitted}
\label{range_bst}
\end{figure}
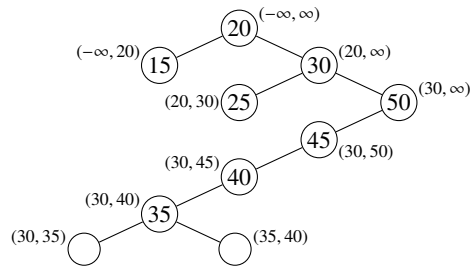

At the leaves, these ranges partition the space of all keys not in the tree. If we reach a leaf node with range $(a, b)$, then we know that keys between $a$ and $b$ 1) are not currently in the tree and 2) can correctly be inserted at this leaf.  For example, suppose we want to insert a key $38$ into the tree in figure~\ref{range_bst}. The right child of the node with key 35 is a leaf with range $(35, 40)$, so it is guaranteed to be the right place to insert 38.

\subsubsection{Per-Node and Global Ghost State}
Importantly, the fact that the leaf ranges partition the space of possible keys remains true even if operations in other threads restructure the tree during our traversal. Furthermore, operations further up the tree can only \emph{increase} the range of nodes below them, so if we find a node whose range includes our key, we are guaranteed to have found the right place for it. This makes ranges ideal for use as ghost state: range information held by one thread will not be invalidated by other threads, and is sufficient to show correctness of BST operations. More precisely, each node's ghost state, identified by a ghost name $g$, contains a pair of its range and its contents, where the contents are $\mathsf{None}$ for a leaf node and $\mathsf{Some}\ (k, v, g_l, g_r)$ for an internal node with key $k$, value $v$, and left and right children with identifiers $g_l$ and $g_r$. The predicate $\nodeghost{g}{(i, j), n}$ asserts that the node with identifier $g$ has range $(i, j)$ and contents $n$, and will be divided between the lock invariant of the node and the abstract state assertion $\treerep$.

The $\treerep$ predicate describes the conditions under which an abstract map $m$ is implemented by the ghost state of a tree. Intuitively, we should be able to gather the per-node ghost states into an abstract tree $t$ that implements the map $m$, in that each key-value pair in $m$ appears in one of the nodes of $t$. We formalize this by first defining the collection of per-node ghost state fragments needed to form an abstract tree:
\[\begin{array}{r l}
& \mathsf{public\_halves}(t, g, (i, j)) \triangleq \mathsf{match}\ t\ \mathsf{with}\\
& \;\qquad |\ \mathsf{Leaf} \Rightarrow \publichalf{g}{(i, j), \mathsf{None}}\\
& \;\qquad |\ \mathsf{Node}\ (k, v, t_l, g_l, t_r, g_r) \Rightarrow \publichalf{g}{(i, j), \mathsf{Some}\ (k, v, g_l, g_r)}\ * \\
& \;\qquad \qquad\mathsf{public\_halves}(t_l, g_l, (i, k)) * \mathsf{public\_halves}(t_r, g_r, (k, j))
\end{array}\]
We recursively walk through the tree, starting from a node $g$ with range $(i, j)$, and collect the $\mathsf{node\_ghost}$ assertions of each node into a single assertion. We compute the range of each node's children as described above: if the key in the current node is $k$, then the left child has range $(i, k)$ and the right child has range $(k, j)$.

This collection of per-node ghost state describes the contents and structure of the tree, but a thread that acquires a node's lock also needs to know that the node actually appears in the tree $t$. For this purpose, we define ghost state assertions $\mathsf{ghost\_nodes}(S)$ and $\mathsf{in\_tree}\ g$, where $S$ is a set of identifiers and $g$ is an identifier, such that $\mathsf{in\_tree}\ g * \mathsf{ghost\_nodes}(S)$ implies that $g \in S$. Then the full definition of $\treerep$ is:
\[\begin{array}{r l}\treerep_g\ m \triangleq &\exists t.\ t \textrm{ implements } m \land \mathsf{public\_halves}(t, g, (-\infty, +\infty))\ * \\&\mathsf{ghost\_nodes}(\mathrm{ids}(t))\end{array}\]
where $\mathrm{ids}(t)$ is the set of all node identifiers in the abstract tree $t$, and $t$ implements $m$ iff the key-value pairs in $t$ are exactly those in $m$. The range of the root node (named $g$) is $(-\infty, +\infty)$, and the $\mathsf{ghost\_nodes}$ assertion guarantees that if any thread or lock holds an assertion $\mathsf{in\_tree}\ g_i$, then $g_i$ is a node in $t$. Thanks to the existential quantification over $t$, we can rearrange its nodes at any time, as long as it remains a well-formed tree with the same set of node IDs and key-value pairs. We will take advantage of this fact when verifying the \lstinline{delete} function.

Next, we connect the per-node locks of our C implementation with the global abstract state of $\treerep$ by associating each node's lock with a lock invariant assertion $\ell \islock R$, where $R$ includes both the concrete points-to assertion for the node and the corresponding abstract $\mathsf{node\_ghost}$. The lock invariant for a node at location $p$ needs to map the abstract contents $c$ to the values present in memory in the \lstinline{t} field of $p$:
\[\begin{array}{l}
\mathsf{match}\ c\ \mathsf{with}\\
|\ \mathsf{None} \Rightarrow p.\texttt{t} = \texttt{NULL}\\
|\ \mathsf{Some}\ (k, v, g_l, g_r) \Rightarrow \exists p_l, p_r.\ p.\texttt{t} \mapsto (k, v, p_l, p_r)
\end{array}\]
We then combine this with the hand-over-hand invariant pattern of section~\ref{selflock}, giving us a final lock invariant of:
\[\begin{array}{r l}
\mathsf{node\_inv}_g(p) \triangleq \ &\exists i\,j\ c.\ \myhalf{g}{(i, j), c} * p.\texttt{lock} \islock_{\pi_1} \mathsf{node\_inv}_g * \\&\mathsf{match}\ c\ \mathsf{with}\\
&|\ \mathsf{None} \Rightarrow p.\texttt{t} = \texttt{NULL}\\
&|\ \mathsf{Some}\ (k, v, g_l, g_r) \Rightarrow k \in (i, j) \land \exists p_l, p_r.\ p.\texttt{t} \mapsto (k, v, p_l, p_r)\ * \\&\qquad p_l.\texttt{lock} \islock_{\pi_2} \mathsf{node\_inv}_{g_l} * p_r.\texttt{lock} \islock_{\pi_2} \mathsf{node\_inv}_{g_r}
\end{array}\]
The hand-over-hand pattern allows us to acquire $p_l.\texttt{lock}$ or $p_r.\texttt{lock}$ and then release $p.\texttt{lock}$ without losing any shares of the lock assertions.

Now we can fill in the pieces of the specifications outlined in section \ref{atomic}:
\begin{mathpar}
\langle m.\,\nodeboxrep_g\,p\ |\ \treerep_g\,m\rangle\ \texttt{insert}(p, k, v)\ \langle \nodeboxrep_g\,p\ |\ \treerep_g\,(m[k \mapsto v])\rangle

\langle m.\,\nodeboxrep_g\,p\ |\ \treerep_g\,m\rangle\ \texttt{lookup}(p, k)\ \langle v.\ \nodeboxrep_g\,p\ |\ \treerep_g\,m \land m(k) = v\rangle

\langle m.\,\nodeboxrep_g\,p\ |\ \treerep_g\,m\rangle\ \texttt{delete}(p, k)\ \langle \nodeboxrep_g\,p\ |\ \treerep_g\,(m[k\mapsto \_])\rangle
\end{mathpar}
where $\treerep_g\,m$ asserts that global ghost state of the tree implements a map $m$, as described above, and $\nodeboxrep$ (the client's handle to the data structure) is simply a reference to the lock for the root node:
\[\nodeboxrep_g\,b \triangleq \exists p.\ b \mapsto p * p.\texttt{lock} \islock \mathsf{node\_inv}_g(p)\]
Next, we will prove that each of our BST functions satisfies its specification.

\subsection{Proofs: Insert and Lookup}

\begin{figure}[ht]
\begin{subfigure}[t]{0.52\textwidth}
\begin{lstlisting}[language = C, basicstyle=\small\ttfamily, numbers=left]
void insert(tree_t** t, int k, void *v){
  tree_t *tgt = *t;
  acquire(tgt->lock);
  while(1) {
    tree* p = tgt->t;
    if (p == NULL) {
      ... // make a new node with k, v
      release(tgt->lock);
      return;
    }
    if (k < p->key){
      void *l_old = tgt->lock;
      tgt = p->left;
      acquire(tgt->lock);
      release(l_old);
    } else if (p->key < k){
      ... // move to p->right instead
    } else {
    	p->value = v;
        release(tgt->lock);
    	return;
    }
  }
} 
\end{lstlisting} 
\end{subfigure}\qquad
\begin{subfigure}[t]{0.4\textwidth}
 \begin{lstlisting}[language = C, basicstyle=\small\ttfamily]
void *lookup(tree_t** t, int k){
  tree_t *tgt = *t;
  acquire(tgt->lock);
  while(1) {
    tree *p = tgt->t;
    if (p == NULL) {

      release(tgt->lock);
      return NULL;
    }
    if (k < p->key){
      void *l_old = tgt->lock;
      tgt = p->left;
      acquire(tgt->lock);
      release(l_old);
    } else if (p->key < k){
      ... // move to p->right instead
    } else {
      void* v = p->value;
      release(tgt->lock);
      return v;
    }
  }
}
\end{lstlisting}
\end{subfigure}
\caption{Code outlines for \lstinline{insert} and \lstinline{lookup}}
\label{bst-conc}
\end{figure}

The code for the insert and lookup functions is shown in figure \ref{bst-conc}. The traversal process is the same in each function: we access the current node's key, compare it to the target key \lstinline{k}, and move to the left child if \lstinline{k} is less than the current key and right if it is greater, using hand-over-hand locking. Thus, both loops have the same invariant. At the beginning of each iteration, we are at a node at pointer $p$ with ghost identifier $g$ and hold the lock for that node, giving us the lock invariant $\mathsf{node\_inv}_g(p)$ with some range $(i, j)$. The loop invariant requires that $\texttt{k} \in (i, j)$, that is, the target key always belongs in the subtree of the current node. This guarantees that if we reach an empty leaf, that leaf is the place where key \lstinline{k} should be held. The actual logic of each operation happens when we either reach an empty leaf node (line 7) or find a node with \lstinline{k} as its key (line 19); there are also the linearization points of the two functions.

When insertion encounters a node with key \lstinline{k}, it simply changes that node's value to \lstinline{value}. If it reaches a leaf, it allocates a new node with key \lstinline{k} and value \lstinline{v}, and inserts it at that position.
In both cases, we must show that we can atomically update the abstract state of the tree from $\treerep\ m$ to $\treerep (m[\texttt{k}\mapsto \texttt{v}])$, by showing that the update to the tree in memory corresponds to a local change in the abstract tree that inserts the new key-value pair. The proofs are split into a proof that the C code implements the local update, and a proof that the local update to a ghost-state tree implements the global update. In the first case, the local change is the insertion of $(\texttt{k}, \texttt{v})$ at an empty leaf whose range includes \lstinline{k}, and the addition of two new empty leaves as its children; in the second case, we simply change the value of an internal node whose key is already \lstinline{k}.

The proof of \lstinline{lookup} is similar, but simpler. We must show that if the key is found in the tree, then the same key-value pair exists in the abstract state, and that if the key is not found then it is not present in the abstract state. The former follows from the lock invariant $\mathsf{node\_inv}$, which guarantees that the key and value in memory match the key and value in the ghost node; the latter follows from the ranges, since if the abstract tree contains a leaf node with range $(i, j)$ such that $\texttt{k} \in (i, j)$, it cannot contain a node with key \lstinline{k}. In either case, the abstract state remains unchanged.

\subsection{Delete}

\begin{figure}[t]
\begin{subfigure}[t]{0.6\textwidth}
\begin{lstlisting}[language = C, basicstyle=\small\ttfamily, numbers=left]
void pushdown_left (tree_t *tgt){
  while(1) {
    tree *p = tgt->t;
    tree_t *rc = p->right;
    void *lq = rc->lock;
    acquire(rc->lock);
    tree *q = rc->t;
    if (q == NULL) {
      tree_t *lc = p->left;
      ... // move lc->t to tgt; deallocate p, lc, and rc
      release(tgt->lock);
      return;
    } else {
      turn_left(tgt, rc);
      tgt = q->left;
      release(tgt->lock);
    }
  }
}
\end{lstlisting} 
\end{subfigure}\qquad
\begin{subfigure}[t]{0.4\textwidth}
 \begin{lstlisting}[language = C, basicstyle=\small\ttfamily]

\end{lstlisting}
\end{subfigure}
\caption{Code outline for \lstinline{pushdown_left}}
\label{pushdown-left}
\end{figure}

The \lstinline{delete} function uses the same traversal logic as \lstinline{insert} and \lstinline{lookup}, but when it finds the node with the target key, it calls a helper function \lstinline{pushdown_left}, outlined in figure~\ref{pushdown-left}. This function rearranges the nodes in the tree, moving the node to be deleted down through the tree until its right child is empty while maintaining the binary search property. This is accomplished by repeatedly rotating sets of three nodes: the target node's right child \lstinline{rc} is moved up to become its parent, with the target node on the left and the right child's right child moved under the target node, as illustrated in figure~\ref{delete-diagram}.
Once the node to be deleted has no right child, its contents are replaced with those of its left child, and its children are removed from the tree.

The \lstinline{pushdown_left} function changes the structure of the tree in two ways. First, the rotation \lstinline{turn_left} (line 14) changes the ranges of the node to be deleted and its right child (e.g., nodes 40 and 55 in the first step of figure~\ref{delete-diagram}); this is a local change, as the ranges of all other nodes (even the children of the rearranged nodes) are preserved. This is not a linearization point, so the rotation must be considered not to change the abstract state of the tree. We accounted for this by quantifying over the abstract tree $t$ in the definition of $\treerep$; rearranging nodes without changing the key-value map still implements the same abstract state, and other threads cannot assume that the tree structure they have observed is still the current structure of the tree. Second, the final removal of the node (line 10) increases the ranges of the subtree below it; for example, in the last step of figure~\ref{delete-diagram}, the range of node 35 changes from $(30, 40)$ to $(30, 50)$ when the node is removed. This is the linearization point of the function. The change occurs while \lstinline{pushdown_left} holds only the lock of the deleted node, so the algebra for range ghost state must allow a node's range to increase at any time, even when the node's lock is held. Fortunately, \lstinline{insert} and \lstinline{lookup} only use ranges to assert that the target key is in the range of the current node, which is still true if the range increases.

\usetikzlibrary{positioning}

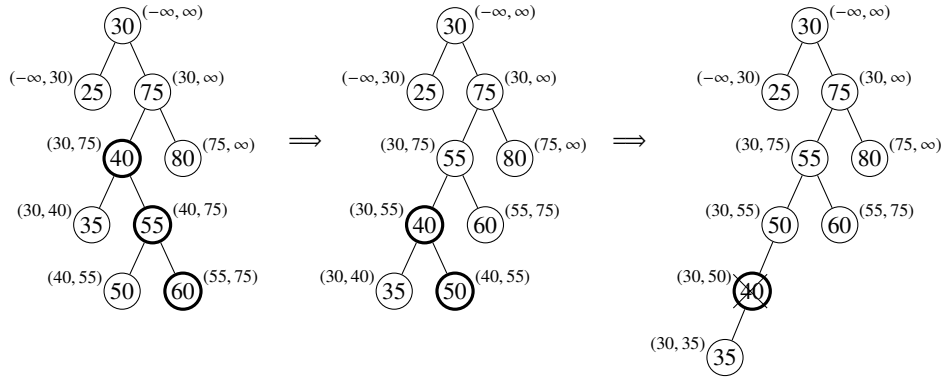
\begin{figure}[t]
\centering
\forestset{
   lab/.style={
        label={[anchor=west, scale=0.75] north east:#1},
    },
}
\begin{forest}
for tree={
    grow=south,
    circle, draw, minimum size=3ex, inner sep=1pt,
    s sep=3mm, l sep=0
        }
    [, phantom, for children={fit=band}, s sep+=20mm,
         before drawing tree={%
          tikz+={%
            \node (a) [inner sep=0pt, fit=(!1) (!1ll) (!1l1ll)] {};
            \node (c) [inner sep=0pt, fit=(!l) (!ll) (!ll111)] {};
            \node (b) [anchor=south] at ($(a.east)!1/2!(c.west)$) {};
            \node [anchor=south] at ($(a.east)!2/5!(b.west)$) {$\Longrightarrow$};
            \node [anchor=south] at ($(b.east)!3/5!(c.west)$) {$\Longrightarrow$};
          },
        },
        [ 30, tikz={\node[right=0pt of .north east, scale=0.75]  {$(-\infty,\infty)$};}
            [25, tikz={\node[left=0pt of .north west, scale=0.75]  {$(-\infty,30)$};}]
            [75, tikz={\node[right=0pt of .north east, scale=0.75]  {$(30,\infty)$};}
                [40, very thick, tikz={\node[left=0pt of .north west, scale=0.75]  {$(30,75)$};}
                    [35, tikz={\node[left=0pt of .north west, scale=0.75]  {$(30,40)$};}]
                    [55, very thick, tikz={\node[right=0pt of .north east, scale=0.75]  {$(40,75)$};}
                        [50, tikz={\node[left=0pt of .north west, scale=0.75]  {$(40,55)$};}]
                        [60, very thick, tikz={\node[right=0pt of .north east, scale=0.75]  {$(55,75)$};}]
                    ]
                ]
                [80, tikz={\node[right=0pt of .north east, scale=0.75]  {$(75,\infty)$};}]
            ]
        ]
        [ 30, tikz={\node[right=0pt of .north east, scale=0.75]  {$(-\infty,\infty)$};}
            [25, tikz={\node[left=0pt of .north west, scale=0.75]  {$(-\infty,30)$};}]
            [75, tikz={\node[right=0pt of .north east, scale=0.75]  {$(30,\infty)$};}
                [55, tikz={\node[left=0pt of .north west, scale=0.75]  {$(30,75)$};}
                    [40, very thick, tikz={\node[left=0pt of .north west, scale=0.75]  {$(30,55)$};}
                        [35, tikz={\node[left=0pt of .north west, scale=0.75]  {$(30,40)$};}]
                        [50, very thick, tikz={\node[right=0pt of .north east, scale=0.75]  {$(40,55)$};}]
                    ]
                    [60, tikz={\node[right=0pt of .north east, scale=0.75]  {$(55,75)$};}]
                ]
                [80, tikz={\node[right=0pt of .north east, scale=0.75]  {$(75,\infty)$};}]
            ]
        ]
        [ 30, tikz={\node[right=0pt of .north east, scale=0.75]  {$(-\infty,\infty)$};}
            [25, tikz={\node[left=0pt of .north west, scale=0.75]  {$(-\infty,30)$};}]
            [75, tikz={\node[right=0pt of .north east, scale=0.75]  {$(30,\infty)$};}
                [55, tikz={\node[left=0pt of .north west, scale=0.75]  {$(30,75)$};}
                    [50, tikz={\node[left=0pt of .north west, scale=0.75]  {$(30,55)$};}
                        [40, very thick, tikz={\node[left=1.8pt of .east, cross out, scale=2, draw]{}; \node[left=0pt of .north west, scale=0.75]  {$(30,50)$};}
                            [35, tikz={\node[left=0pt of .north west, scale=0.75]  {$(30,35)$};}]
                            [,no edge, draw=none]
                        ]
                        [,no edge, draw=none]
                    ]
                    [60, tikz={\node[right=0pt of .north east, scale=0.75]  {$(55,75)$};}]
                ]
                [80, tikz={\node[right=0pt of .north east, scale=0.75]  {$(75,\infty)$};}]
            ]
        ]
        \ignore{[ 30, tikz={\node[right=0pt of .north east, scale=0.75]  {$(-\infty,\infty)$};}
            [25, tikz={\node[left=0pt of .north west, scale=0.75]  {$(-\infty,30)$};}]
            [75, tikz={\node[right=0pt of .north east, scale=0.75]  {$(30,\infty)$};}
                [55, tikz={\node[left=0pt of .north west, scale=0.75]  {$(30,75)$};}
                    [50, tikz={\node[left=0pt of .north west, scale=0.75]  {$(30,55)$};}
                        [35, tikz={\node[left=0pt of .north west, scale=0.75]  {$(30,40)$};}]
                        [,no edge, draw=none]
                    ]
                    [60, tikz={\node[right=0pt of .north east, scale=0.75]  {$(55,75)$};}]
                ]
                [80, tikz={\node[right=0pt of .north east, scale=0.75]  {$(75,\infty)$};}]
            ]
        ]}
    ]
\end{forest}
\caption{Deleting node 40 from the tree via \lstinline{pushdown_left}}
\label{delete-diagram}
\end{figure}

\section{Using the Specifications}
To demonstrate that the complexities of our lock-based proofs are hidden from prospective users, we verified a simple client program for the binary search tree, shown in figure \ref{bst-conc-client}. A producer thread running the \lstinline{update_tree} function populates a tree, while a consumer thread running \lstinline{retrieve_value} waits until key 2 is associated with a pointer to the value 4, and then returns the value associated with key 1. This is a data-structure-level version of the message-passing idiom, and demonstrates one of the advantages of atomic specifications: we can use them to synchronize threads and pass ownership of resources via data structure operations. The invariant for the BST \lstinline{t} is a simple state machine with three states, depending on the contents of the abstract map $m$: 1) $m(2) = \texttt{NULL}$, 2) $m(2) \mapsto i$ for some integer $i \neq 4$; and 3) $m(2) \mapsto 4 \land m(1) \mapsto 3$. In this final state, the invariant also holds the resources to be transferred from the producer to the consumer. We also include ghost state to reflect the fact that there is exactly one producer thread (so the producer always knows exactly which values are in the tree) and one consumer thread (so it can retrieve the resources without worrying about interference). The proof of the client uses only the atomic specifications of the BST operations, and is not affected by the style of lock specification.

\begin{figure}[htb]
\begin{subfigure}[t]{0.4\textwidth}
\begin{lstlisting}[language = C, basicstyle=\small\ttfamily, numbers=left]
void* update_tree(void* t){
  a = 1;
  b = 2;
  c = 3; 
  d = 4;
  insert(t, 1, &a);
  insert(t, 2, &b);
  insert(t, 1, &c);
  insert(t, 2, &d);
  return NULL;
}
\end{lstlisting} 
\end{subfigure}\qquad
\begin{subfigure}[t]{0.55\textwidth}
 \begin{lstlisting}[language = C, basicstyle=\small\ttfamily]
int retrieve_value(treebox t){
  int y = -1;
  do {
     searchResult = lookup (t,  2);
     if (searchResult != NULL) {
        y = *((int *) searchResult);
     }
  } while (y != 4);
  void* searchResult = lookup (t,  1);
  return *((int *) searchResult);
}
\end{lstlisting}
\end{subfigure}
\caption{Client code}
\label{bst-conc-client}
\end{figure}


\section{Related Work}
\label{related}
Gotsman et al.~\cite{gotsman}, in the same paper in which they introduced invariant-style lock specs, also verified a linked list with hand-over-hand locking, which became a common example for CSLs that handled fine-grained locking. Among the most recent of these is the work of Krishna et al.~\cite{templates}, who used atomic lock specs to prove correctness of the same data structure. This work updates the work of Gotsman et al. with modern specifications, and directly compares the results to those of Krishna et al.

VeriFast~\cite{verifast}, a separation-logic-based verifier for C and
Java, supports lock-based and atomic
concurrency~\cite{verifast-conc}, and has been used to verify a
hand-over-hand-locking linked list similar to that of Krishna et al.
The specifications of the lock operations and the list itself use
a precursor of TaDA's logical atomicity. VeriFast is not 
foundational, but its basic logic is verified
against the semantics of a toy language in Coq.

\section{Conclusion}
\label{conclusion}
While most new CSL proofs of lock-based data structures use TaDA-style atomic specs for their locks, the implications of the switch away from invariant-based specs had not been thoroughly examined. We have now demonstrated that invariant-based specs can still be used to prove logically atomic specs for larger data structures (even for fine-grained implementations) and that this requires somewhat more complex proofs than using atomic lock specs (especially for fine-grained implementations). The old style of lock spec may still be used in systems like VST, where switching would invalidate existing automation or soundness proofs, and in these systems the approach we have outlined can be used to obtain strong correctness properties for fine-grained-locking data structures. When possible, however, we would prefer to use atomic specifications for locks to simplify our proofs. Indeed, we intend to investigate shifting the foundations of VST to use atomic operations and derive atomic specs for locks: while updating the soundness proof will be a considerable effort, using the newer style of lock specification will simplify all future proofs about lock-based data structures.

\subsubsection{Acknowledgements} This work benefited greatly from correspondence with Ralf Jung, who pointed out Zhang's syncer approach and explained the finer points of logical atomicity in Iris.

\bibliography{sources}
\end{document}

%% file: intro.tex


Concurrent separation logic (CSL)~\cite{csl} is a useful tool for proving correctness of concurrent programs. While early iterations focused on data structures implemented with locks and specifications involving memory safety and data-race-freedom~\cite{gotsman,oraclesematic}, more recent logics are able to prove linearizability-style functional correctness of lock-free implementations that use low-level atomic memory operations~\cite{tada,iris}, even under weak memory models~\cite{rustbelt-relaxed}. In particular, \emph{logical atomicity}~\cite{tada} has become the gold standard for concurrent data structure specifications, capturing the idea that data structure operations should appear to take place instantaneously with no externally visible intermediate states. The effects of a logically atomic operation become visible at a \emph{linearization point}, a single instruction (usually an atomic memory operation or call to a logically atomic function) that publishes the new state of the data structure.

Intuitively, a critical section in a lock-based implementation serves the same role as an atomic instruction in a lock-free data structure, and it is not hard to identify linearization points in lock-based implementations---the effects of an update become visible when the lock protecting the updated data is released. The translation of this intuition into proof, however, depends on the specifications used for lock operations. The CSL literature contains two distinct specifications for lock acquire and release:
\begin{enumerate}
\item In Gotsman/Hobor-style specs~\cite{gotsman,oraclesematic}, each lock is associated with a lock invariant $R$, which is gained by \texttt{acquire} operations and restored by \texttt{release} operations.
\item In TaDA-style specs~\cite{tada}, \texttt{acquire} atomically moves a lock from the unlocked state to the locked state, and \texttt{release} does the reverse.
\end{enumerate}
Style 2 specs are known to imply style 1 specs, and have been the style of choice for most recent verifications~\cite{tada-live,templates}. However, unpublished work by Zhang~\cite{atomic-syncer} shows that style 1 can also be used to prove atomic specifications for data structures.

In this paper, we adapt this approach to the Verified Software Toolchain~\cite{plfcc}, a system for proving correctness of C programs that has recently been extended with support for general ghost state and atomicity proofs. The soundness proof for VST~\cite{cpm} lifts the single-threaded correctness theorem of the CompCert compiler~\cite{compcert} to a concurrent setting, and relies explicitly on style 1 lock specs; modifying this proof to rely on atomic operations instead is both theoretically and practically daunting. We demonstrate that 1) \textbf{our approach can be used to prove atomic specs for interesting data structures with fine-grained locking, starting from style 1 lock specs}, and 2) \textbf{these proofs require some additional complications that can be avoided with style 2 lock specs}. The key difference is that in style 1 a thread must gain ownership of a handle to a component's lock before acquiring the lock and reading/modifying the component, while in style 2 the locks and components alike can be considered part of a single abstract state that is accessed at each lock access. However, the top-level specifications proved for the data structure are not affected by the complexity of the lock specs, so proofs of clients of the data structure are no more complex than they would be otherwise.

To our knowledge, this is the first formal comparison of the two styles of lock specifications, and our technique should be useful in tools like VST where the older style is deeply integrated into the tool. This is also the first \emph{foundational} verification of a C implementation against logically atomic specifications; prior tools such as VeriFast~\cite{verifast-conc} have been used to verify C implementations, but are not proved sound against a C semantics and so can avoid the complexities of lock specs w.r.t. concurrent soundness.